# Direct Imaging of Complex Spin Ice Behavior and Ordered Sublattices in Artificial Ferromagnetic Quasicrystals

V. S. Bhat[1], A. Balk[2,3], B. Farmer[1], E. Teipel[1], N. Smith[1], J. Unguris[2], J. T. Hastings[4] and L. E. De Long[1]

[1]*University of Kentucky, Department of Physics and Astronomy*
*505 Rose Street, Lexington, KY, USA 40506-0055*

[2]*National Institute of Standards and Technology*
*Center for Nanoscale Science and Technology*
*100 Bureau Dr., Gaithersburg, MD, USA 20899*

[3]*Maryland Nanocenter, University of Maryland, College Park, MD, USA 20742*

[4]*University of Kentucky, Department of Electrical and Computer Engineering*
*453F Paul Anderson Tower, Lexington, KY, USA 40506-0046*

We have imaged magnetization textures of permalloy films patterned into Penrose P2 tilings (P2T) using scanning electron microscopy with polarization analysis (SEMPA). P2T film segments have near-uniform, bipolar magnetization, similar to artificial spin ices, but with asymmetric vertex coordination that induces a more "complex spin ice" behavior mediated by exchange interactions in vertex domain walls. Numerical simulations including long-range dipole interactions agree with SEMPA images of as-grown P2T, and predict a ferromagnetic ground state for a 2D P2T lattice of classical Ising spins.

Bulk quasicrystals are rarely found in nature, are difficult to grow and characterize in the laboratory, but exhibit unique physical properties [**1-3**]. Their signature long-range structural order without periodic translational symmetry places them in a unique niche between periodic crystals and amorphous materials [**4-6**]. In particular, magnetic interactions in known bulk quasicrystals result in spin glass, rather than long-range magnetic order [**1,7,8**]. Fortunately, advances in nanofabrication permit ferromagnetic thin films to be patterned into "artificial quasicrystals" (see Fig. 1) whose complex dynamics and magnetic reversal can be systematically controlled via tiling design [**4,9**]. However, the effects of quasicrystal symmetry on cooperative magnetic behavior and attainment of an equilibrium groundstate have not been experimentally investigated for this novel class of metamaterials.

SEMPA [**10**] is a superior tool for direct characterization of thin-film magnetization on a sub-micrometer scale, including domain wall (DW) structures within vertices connecting segments [**11**]. Secondary electrons ejected from a magnetic sample maintain their spin polarization, which is detected and correlated with the position of the excitation electron beam in SEMPA. Simultaneous sensitivity to two orthogonal, in-plane directions yields maps of net in-plane polarizations. The sample surface is ion-milled ($\approx$ 1 keV Ar beam) to remove surface contaminants. One or two monolayers of Fe are then deposited *in situ*, to enhance domain contrast. The resulting magnetic images are masked with simultaneously obtained secondary electron images to highlight magnetic regions over surrounding non-magnetic areas (see Figs. 2a, 2b).

We generated P2T via the "deflation method" using a graphics algorithm [**12-14**] incorporated into our electron beam lithography software. P2T samples were patterned using positive resist spun onto Si substrates, followed by slow (0.01 nm/s) permalloy deposition using electron beam evaporation and resist lift-off. Each P2T was truncated to a "3$^{rd}$ generation" decahedron of 6.8 μm apical width; and a 16 × 16 square array of P2T decahedra with 11 μm spacing was fabricated without unpatterned film on substrate borders. A SEM image shown in Fig. 1 shows elongated film segments of two lengths $d_{1,2}$ and corresponding magnetic moments $m_{1,2} \propto d_{1,2}$, connected by asymmetric vertices with a range of coordination numbers $2 \leq N \leq 5$ (see Supplemental Information [**SI**], Figs. SI1-SI3).

The high length/width ratio $d/w \geq$ (500 nm/70 nm) $\approx$ 7 favors uniform, bipolar (Ising-like) magnetizations, whereas our SEMPA images reveal magnetizations imperfectly aligned with long segment axes, as shown in Figs. 2a and 2b. Nevertheless, coarse-grained averages of P2T segment magnetizations approximate macroscopic

Ising spins (see Figs. 2c and 2d), similar to artificial spin ices [15-17] (ASI).

However, most ASI studied to date are periodic arrays of identical Ising segments with symmetric vertices of one, low coordination value, which obey "spin ice rules" (SIR) [15-18]. The stability of ASI can be analyzed using a "dumbbell" model originally developed [19] for *microscopic* spin ice: The Ising spins are replaced by magnetic charges of equal magnitude and opposite sign placed on the ends of rigid rods. Nearest-neighbor exchange and longer-range dipolar interaction energies are then calculated using a Coulomb model that requires net vertex charge to be minimized [17,20].

However, multiply-connected P2T vertices harbor DW whose energies are neglected in dumbbell models that consider only dipole interactions in periodic ASI [17]. Alternatively, we use the (zero-temperature) Object Oriented MicroMagnetic Framework [21] (OOMMF) to simulate nearest-neighbor dipole and short-range exchange interactions that create and pin DW in vertices. A key to our approach is to treat segments intersecting at a given vertex as isolated "vertex clusters" classified by angular asymmetry. A digital image of an ideal P2T is imported using a 5 nm x 5 nm × 25 nm grid. Permalloy saturation magnetization $M_S = 8.6 \times 10^5$ A·m$^{-1}$ and exchange stiffness $A = 13 \times 10^{-12}$ J·m$^{-1}$ are used to calculate magnetostatic energies in zero field for each of the $2^N$ possible configurations of segment polarizations for the nine, *N*-fold cluster types (see Fig. 3).

For example, OOMMF simulations favor a "soft" DW configuration for which the magnetization "bends through" the vertex ("low-energy" case in Fig. 3c), over a "hard", head-to-head or tail-to-tail DW ("high-energy" case in Fig. 3b). Although both soft and hard vertex DW configurations obey SIR, a corresponding dumbbell model calculation disagrees by yielding a vertex energy for Fig. 3c *greater* than that of Fig. 3b [**SI**].

We compared the simulated vertex cluster energies with SEMPA images of ten as-grown P2T samples (each with 121 vertices), and found 104 high-energy vertices (44 of which disobey SIR), which is only 8.6 % (3.6 %) of the total population (1210) of vertices sampled. The zero-temperature cluster simulations are therefore in satisfactory agreement with SEMPA images, and suggest the imaged *high-energy vertices are dilute topological defects quenched into an ordered ground state*. To verify this conjecture, we summed the individual vertex cluster energies to approximate the total P2T magnetostatic energy, which was minimized using a finite-temperature ("simulated annealing" [22]) Monte Carlo (MC) algorithm [**SI**].

Our MC cluster analysis demonstrates the preference for "soft" DW in asymmetric P2T vertices (Fig. 3) *favors ferromagnetic coupling* between adjacent segments and imposes "tangential" orientations on two highly correlated sublattices SL3 and SL4, defined in Fig. 4a (see Fig. SI6 [**SI**]). The SEMPA image of as-grown Sample 8-2 in Fig. 2a confirms SL3 segment polarizations self-organize into a clockwise loop on the P2T central "star" (Fig. 2a inset). The larger dipole loops present in SL4 are made more apparent in Figs. 4a and 4b.

Although strong nearest-neighbor interactions fix SL3 and SL4 segments in ordered, low-cluster-energy states, the polarizations of other segments *remain frustrated*, as shown in Fig. 4c: For example, consider CN5-1 vertices (yellow "wheels") that are connected to SL4 vertices. SL4 ordering does not constrain the net polarizations (white arrows) of the five CN5-1 clusters, therefore each CN5-1 cluster has a tenfold cluster-energy degeneracy. A lower limit for the total cluster energy degeneracy of the P2T state shown in Fig. 4c, considering only the five CN5-1 vertices, is $1.0 \times 10^5$. Consideration of other weakly correlated SL1 and SL2 vertices yields a value of $3.28 \times 10^{10}$ for the lower degeneracy limit [**SI**].

A high degeneracy in total vertex cluster energy implies *long-range dipole correlations are decisive in fully ordering a P2T ground state*. We therefore performed higher-resolution MC simulations made tractable by holding strongly correlated SL3 and SL4 segments in their persistently ordered states (as observed in both SEMPA images and MC simulations), while permitting weakly correlated segments to switch among configurations with low cluster energies. Inclusion of all dipolar interactions between P2T segments was sufficient to drive the weakly correlated segments into two additional ordered sublattices classified according to their symmetries (see Fig. 4a): SL1 exhibited no net moment and fivefold rotational symmetry, and SL2 exhibited mirror symmetry about a net ferromagnetic moment. The four ordered sublattices identified by MC comprise a *magnetically ordered P2T ground state* shown in Fig. 4a.

Attainment of the FM ground state predicted by MC is problematic, since micrometer-scale film segments introduce energy barriers $\approx 10^5$ K, which suppresses thermal fluctuations of segment magnetizations and impedes their reversal at room temperature [23]. ASI samples therefore retain non-equilibrium, quenched magnetization textures during film deposition [17]. Consequently, we employed a low permalloy deposition rate (0.01 nm·s$^{-1}$) to allow access to a wider variety of microstates before the segment blocking temperature is reached [24]. Partial equilibration is evident in a SEMPA image showing two (time-reversed) superdomains coexisting on SL4 in the as-grown state (see Fig. 4b). We note all high-energy vertices lie on walls between superdomains, which is similar to the low-energy textures of periodic ASI [25]. All (104) high-energy vertices observed in SEMPA images of our ten as-grown P2T samples are located in SL4 (see Fig. SI7 [**SI**]).

Nevertheless, the P2T ground state of Fig. 4a is very different from those proposed for disconnected, periodic, honeycomb [20] and square [26] ASI, which are governed only by dipolar coupling between segments. The apparent tendency to form large dipole loops in as-grown P2T originates in the exchange energy cost to form asymmetric, hard vertex DW (Fig. 3); this contrasts with periodic ASI, where dipole interactions favor a ground state with small ordered loops ("vortices") of near-neighbor segments [20,26,27]. In the case of P2T, large numbers of vortices are found in simulations of field-cycled samples in applied fields near mid-reversal [4], which is a field regime inaccessible to our SEMPA. Such vortices are absent in SEMPA images of the field-cycled remnant state of Sample 8-2 (Figs. 2b, 2d); the high- and low-energy vertices and a high degree of mirror symmetry with respect to the magnetic field axis apparent in Figs. 2b and 2d are anticipated for a "near-saturated" Ising state [9] in the presence of energy barriers to equilibration. In contrast, SEMPA images of as-grown P2T exhibit low-energy defects with vertices that obey SIR, but lack mirror symmetry by favoring large closed loops of Ising segments. The striking differences between as-grown and field-cycled remnant states (see Figs. SI7, SI8 and Table SI1) confirm our slow film deposition protocol enhances equilibration of as-grown samples.

In conclusion, we have used SEMPA to acquire the first direct, two-dimensional images of magnetization textures of ferromagnetic films patterned into artificial quasicrystals that constitute a new class of *complex artificial spin ices*, whose magnetic textures are strikingly different from those of previously studied *periodic* ASI. The aperiodic, global five-fold symmetry of P2T requires several asymmetric coordinations of multiply connected vertices with DW that are the primary drivers of complex spin ice behavior. A high degree of dipole order in sublattices strongly correlated by nearest-neighbor exchange and dipole interactions are mixed with frustrated sublattices weakly correlated by longer-range dipole interactions. Moreover, our Monte Carlo simulations have identified an emergent composite of perfectly ordered P2T sublattices as the ordered ferromagnetic ground state of an artificial quasicrystal of classical Ising spins, which contrasts the persistent non-observation of long-range magnetic order in bulk quasicrystals [5-8].

Finally, our results suggest that the variety of vertex coordinations and symmetries exhibited by known quasicrystalline tilings [1-3,12,13] will provide a rich set of metamaterials for highly-controlled experimental and numerical studies of effects of frustration and aperiodicity on "quasimagnetic order" [5,6] and complex spin ice behavior. Our SEMPA images strongly suggest our MC ground state will be directly observed in future studies of better equilibrated P2T.

Research at University of Kentucky supported by U.S. DoE Grant DE-FG02-97ER45653, the UK Centers for Advanced Materials, Computational Sciences, and Nanoscale Science and Engineering. AB acknowledges support of this research under the Cooperative Research Agreement between the University of Maryland and National Institute of Standards and Technology Center for Nanoscale Science and Technology, Award 70NANB10H193, through the University of Maryland. J. P. Straley, K. Ross and F. Guo contributed helpful criticism.

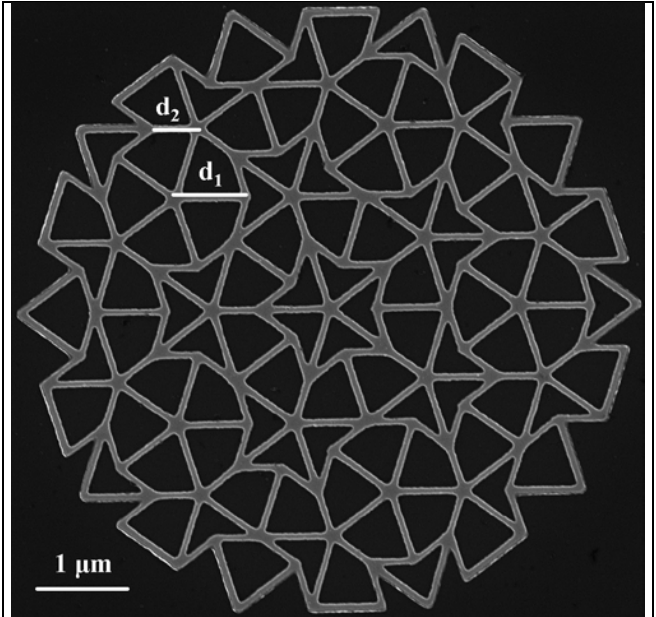

FIG. 1. Bright (dark) regions correspond to permalloy film (Si substrate) of magnetization $M$. Long ($d_1$) and short ($d_2$) segment lengths were nominally 810 ± 10 nm and 500 ± 10 nm, respectively. The width $w$ and thickness $t$ of Permalloy segments were nominally 70 ± 10 nm and 25± 1 nm, respectively. The dumbbell magnetic charge $q = Mwt$ associated with a segment end is independent of segment length $d$. Total apical width of the pattern is 6.8 μm. Note four distinct configurations of $N = 5$ vertices, a single $N = 4$ vertex, two $N = 3$ vertices (one on an edge), and two $N = 2$ edge vertices (see Fig. 3a).

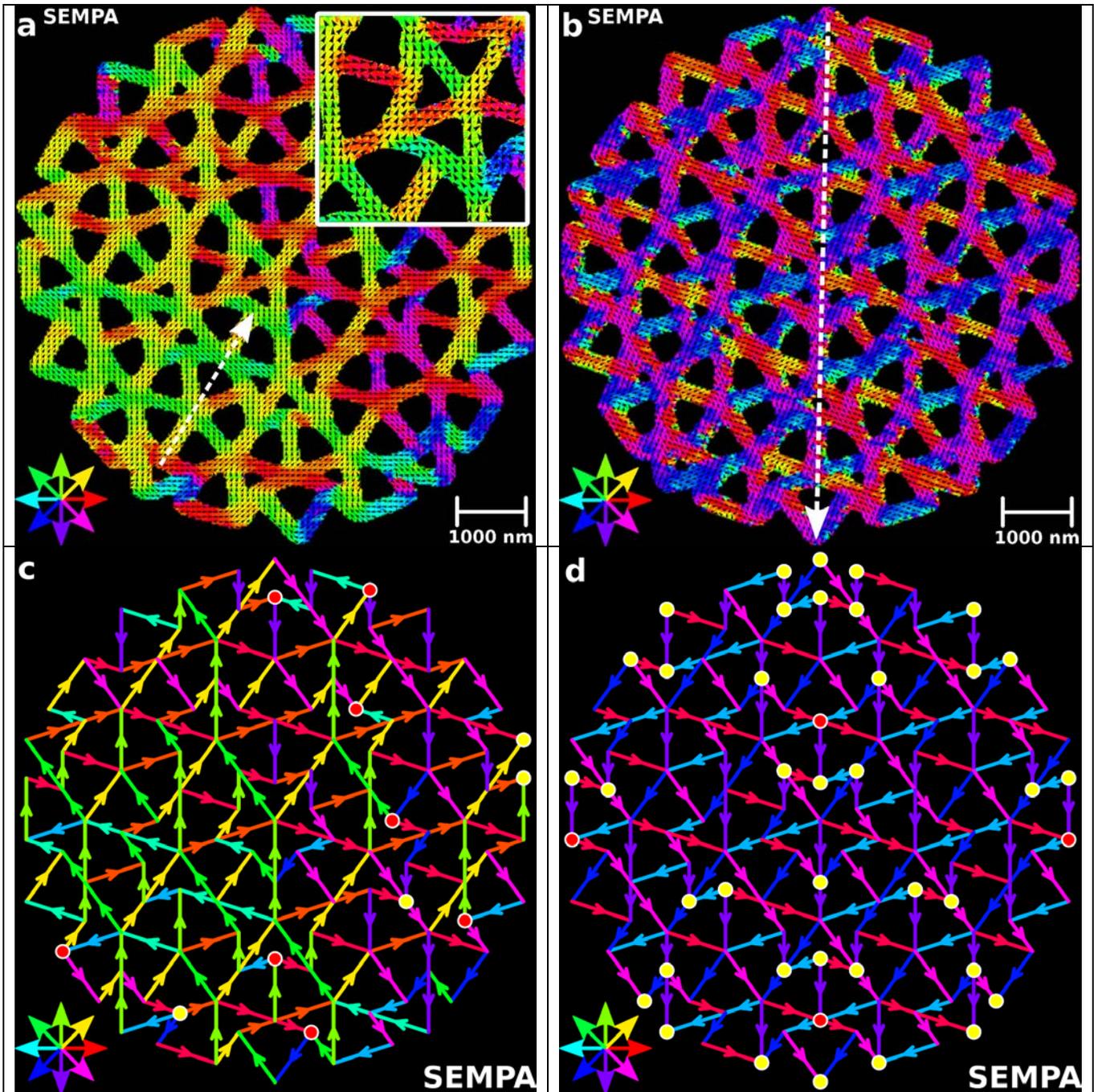

FIG. 2. Color compasses denote in-plane magnetization direction. (a) SEMPA image of as-grown state. White arrow denotes a flux closure loop extending around central P2T "star". Inset: High-resolution SEMPA image showing central star vertex DW for a low-energy, 2-in/3-out segment configuration. Black arrows indicate in-plane magnetization direction. (b) SEMPA image of remnant state after applied field protocol, $H = 0$ A·m$^{-1}$ → +7.96 x 10$^4$ A·m$^{-1}$ → -7.96 x 10$^4$ A·m$^{-1}$ → +7.96 x 10$^4$ A·m$^{-1}$ → 0 A·m$^{-1}$, along direction of white dotted arrow. Note clear differences in texture between (a) and (b). (c) Dipole map of (a) showing coarse-grained (Ising) polarizations of segments denoted by colored arrows. Red dots indicate high-energy vertices that obey SIR. Yellow dots indicate higher-energy vertices that violate SIR. (d) Dipole map of (b) exhibiting many violations of SIR (yellow dots). Note *perfect mirror symmetry* of magnetization about applied field direction in (b) and (d).

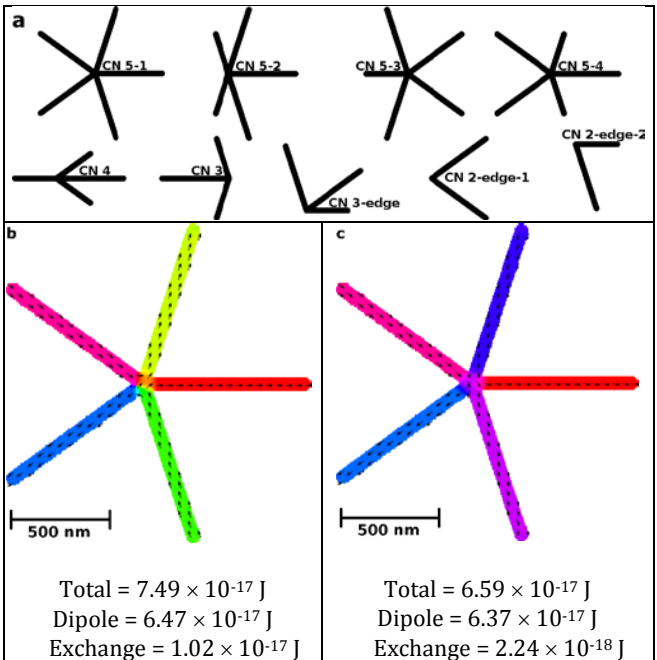

| Total = 7.49 × 10$^{-17}$ J | Total = 6.59 × 10$^{-17}$ J |
| Dipole = 6.47 × 10$^{-17}$ J | Dipole = 6.37 × 10$^{-17}$ J |
| Exchange = 1.02 × 10$^{-17}$ J | Exchange = 2.24 × 10$^{-18}$ J |

FIG. 3. (a) (Top) Nine possible vertex cluster configurations labeled according to asymmetric coordination for third-generation P2T. There are four distinct cluster energies for the symmetric CN5-1 vertex. The highest energy is attained for all-5-in or 5-out segment polarizations. The next-highest is 4-in/1-out or 1-in/4-out. The lowest is 3-in/2-out or 2-in/3-out configurations obeying SIR. OOMMF cluster simulations indicate a 1.6 % difference in the dipole energies (listed above), and a 127 % difference in the exchange energies between a higher-energy, 2-in/3-out configuration (b) and a lower-energy, 2-in/3-out configuration (c). The total cluster energy difference is due to creation of hard, "head-to-head" or "tail-to-tail" DW in vertex (b), versus smoothly "bending" textures in vertex (c) (Note the dumbbell model energy predicts the opposite order of stability [**SI**]).

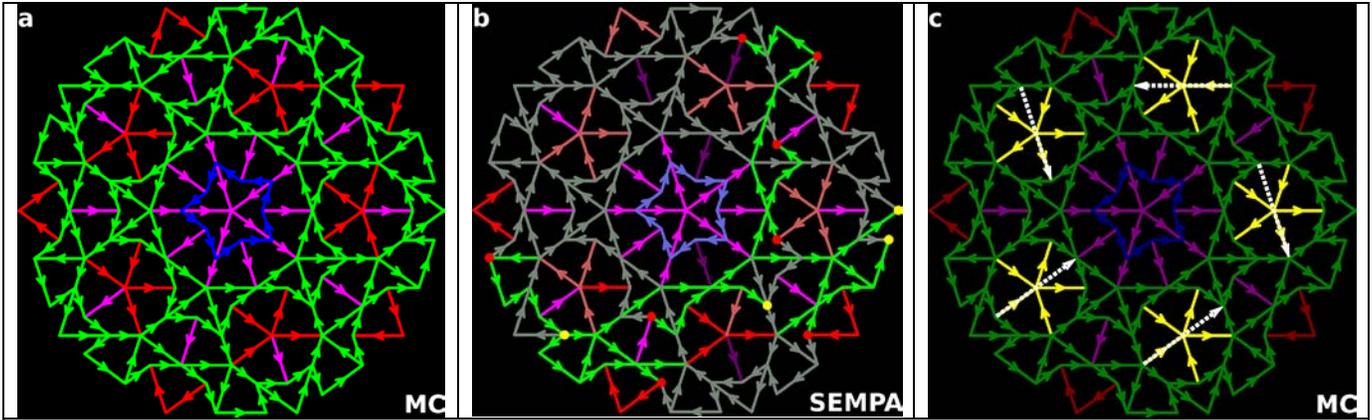

FIG. 4. (a) Ordered MC ground state obtained by including long-range dipole interactions. SL1 (red) dipoles exhibit 5-fold rotational symmetry and no net moment. SL2 (magenta) dipoles exhibit mirror symmetry about a net ferromagnetic axis. SL3 (dark-blue) and SL4 (green) dipoles are correlated by cluster interactions [**SI**]. (b) Sublattice map for SEMPA image of as-grown Sample 8-2 (shown in Fig. 2a and Fig. 2c). An ordered SL3 is shown in light blue. Two ordered superdomains of SL4 are colored grey (clockwise vorticity) and lime-green (counter-clockwise vorticity), respectively. Salmon dipoles are disordered SL1 segments, and dark magenta dipoles are disordered SL2 segments. Superdomains are separated by red (yellow) dots that are high-energy vertices obeying (disobeying) SIR. Low-energy vertices obeying SIR are not highlighted. (c) MC ground state of (a) altered by substitution of one of 10 configurations of the CN5-1 segments (highlighted in yellow) with degenerate total cluster energies. The SL1 and SL2 segment polarizations in CN5-1 are not affected by SL4 order in cluster energy MC; long-range dipole interactions remove cluster degeneracies and stabilize the ground state in (a). White arrows denote (disordered) net cluster moment directions.